\documentclass[aps,prb,twocolumn,superscriptaddress,showpacs,amsmath,amssymb]{revtex4}
\usepackage{graphicx}

\begin{document}

\title{Local density of states and superconducting gap in the iron chalcogenide superconductor Fe$_{1+\delta}$Se$_{1-x}$Te$_{x}$ observed by scanning tunneling spectroscopy}

\author{Takuya~Kato}
\email[]{tkato@rs.kagu.tus.ac.jp}
\affiliation{Department of Physics, Tokyo University of Science, 1-3 Kagurazaka, Shinjuku-ku, Tokyo 162-8601, Japan}

\author{Yoshikazu~Mizuguchi}
\affiliation{Superconducting Materials Center, National Institute for Materials Science, 1-2-1 Sengen, Tsukuba, Ibaraki 305-0047, Japan}
\affiliation{Japan Science and Technology Agency--Transformative Research-Project on Iron-Pnictides (JST--TRIP), 1-2-1 Sengen, Tsukuba, Ibaraki 305-0047, Japan}
\affiliation{Graduate School of Pure and Applied Sciences, University of Tsukuba, 1-1-1 Tennodai, Tsukuba, Ibaraki 305-8577, Japan}

\author{Hiroshi~Nakamura}
\affiliation{Department of Physics, Tokyo University of Science, 1-3 Kagurazaka, Shinjuku-ku, Tokyo 162-8601, Japan}

\author{Tadashi~Machida}
\affiliation{Department of Physics, Tokyo University of Science, 1-3 Kagurazaka, Shinjuku-ku, Tokyo 162-8601, Japan}
\affiliation{Superconducting Materials Center, National Institute for Materials Science, 1-2-1 Sengen, Tsukuba, Ibaraki 305-0047, Japan}

\author{Hideaki~Sakata}
\affiliation{Department of Physics, Tokyo University of Science, 1-3 Kagurazaka, Shinjuku-ku, Tokyo 162-8601, Japan}

\author{Yoshihiko~Takano}
\affiliation{Superconducting Materials Center, National Institute for Materials Science, 1-2-1 Sengen, Tsukuba, Ibaraki 305-0047, Japan}
\affiliation{Japan Science and Technology Agency--Transformative Research-Project on Iron-Pnictides (JST--TRIP), 1-2-1 Sengen, Tsukuba, Ibaraki 305-0047, Japan}
\affiliation{Graduate School of Pure and Applied Sciences, University of Tsukuba, 1-1-1 Tennodai, Tsukuba, Ibaraki 305-8577, Japan}

\date{\today}

\begin{abstract}
We report on the investigation of the quasiparticle local density of states and superconducting gap in the iron chalcogenide superconductor Fe$_{1+\delta}$Se$_{1-x}$Te$_{x}$ ($T_{\mathrm{c}} \sim 14$~K).
The surface of a cleaved crystal revealed an atomic square lattice, superimposed on the inhomogeneous background, with a lattice constant of $\sim 3.8$~\AA\ without any reconstruction.
Tunneling spectra measured at 4.2~K exhibit the superconducting gap, which completely disappears at 18~K, with a magnitude of $\sim 2.3$~meV, corresponding to $2\Delta / k_{\mathrm{B}}T_{\mathrm{c}}=3.8$.
In stark contrast to the cuprate superconductors, the value of the observed superconducting gap is relatively homogeneous, following a sharp distribution with a small standard deviation of 0.23~meV.
Conversely, the normal-state local density of states observed above $T_{\mathrm{c}}$ shows spatial variation over a wide energy range of more than 1~eV, probably due to the excess iron present in the crystal.
\end{abstract}


\maketitle

Since the discovery of superconductivity in LaFeAsO$_{1-x}$F$_{x}$ (the so-called 1111 system) at 26~K, several other iron-based high-$T_{\mathrm{c}}$ superconductors with similar layered structures have been found, such as $Ae$Fe$_2$As$_2$ (the 122 system, where $Ae$ is an alkaline-earth metal) and Fe$Ch$ (the 11 system, where $Ch$ is a chalcogenide).\cite{kamihara:jacs2008,rotter:prl2008,wang:ssc2008,hsu:pnas2008,mizuguchi:apl2009,ogino:sst2009}
Among these materials, the iron chalcogenide superconductor FeSe ($T_{\mathrm{c}} \sim 8$~K) has attracted much attention as the simplest model system.
By partially substituting Te for Se, $T_{\mathrm{c}}$ can be enhanced up to 14~K.\cite{yeh:epl2008,fang:prb2008}
Moreover, the application of pressure increases the $T_{\mathrm{c}}$ of FeSe to 37~K.\cite{mizuguchi:apl2008,margadonna:prb2009,medvedev:natmat2009,masaki:jpsj2009,imai:prl2009}
This dramatic effect is likely to be related to the enhancement of magnetic fluctuations under pressure.
In order to understand the mechanism of the superconductivity in these iron-based compounds, the identification of the superconducting gap is necessary.
Scanning tunneling spectroscopy (STS) is one of the most powerful tools for providing information on the superconducting gap, because it is capable of measuring the quasiparticle local density of states (LDOS) with atomic resolution.
This ability to investigate the electronic states in real space has resulted in STS studies being intensively performed on the cuprate high-$T_{\mathrm{c}}$ superconductors.
Many important aspects of cuprate superconductivity have been revealed using STS such as the observation of the superconducting gap and pseudogap,\cite{renner:prl1998} the nanoscale inhomogeneous nature of the quasiparticle excitation,\cite{pan:nature2001,kato:jpsj2008} checkerboard like electronic order,\cite{hangauri:nature2004,machida:jpsj2006} impurity states,\cite{pan:nature2000} and vortex states.\cite{pan:prl2000,matsuba:jpsj2007}
Following the discovery of superconductivity in the iron-based compounds, analogous STS studies have been carried out.\cite{millo:prb2008,boyer:08064400,pan:08080895,yin:prl2009,massee:prb2009,yin:physicac2009}
The superconducting gap in the LDOS has consequently been observed in the 1111 and 122 systems, but the cleaved surfaces of these materials exhibit a complex structure that depends on the nature of the exposed topmost plane and the cleavage temperature.
In contrast, the 11 system is more suited to STS studies because there is a single cleavage plane, the iron chalcogenide layer, which is free from reconstruction and unaffected by the cleavage conditions.\cite{massee:09075544}
Furthermore, the investigation of the LDOS and superconducting gap in the simplest systems of the iron-based compounds is crucial in order to understand the mechanisms of superconductivity.
In this letter, we report on the observation of the LDOS and superconducting gap in the iron chalcogenide superconductor Fe$_{1+\delta}$Se$_{1-x}$Te$_{x}$.

Single crystals of Fe$_{1+\delta}$Se$_{1-x}$Te$_{x}$ were grown using a self-flux method.
A mixture of Fe, Se, and Te powders with a starting composition of FeSe$_{0.25}$Te$_{0.75}$ was placed in an alumina crucible, which was then sealed in an evacuated quartz tube.
The tube was heated at 950$^\circ\mathrm{C}$ for 10~h, cooled to 700$^\circ\mathrm{C}$ at a rate of $-5^\circ\mathrm{C}/\mathrm{h}$, and then annealed at 400$^\circ\mathrm{C}$ for 100~h.
The composition of the crystal obtained was determined as Fe$_{1.05}$Se$_{0.15}$Te$_{0.85}$ using the energy dispersive x-ray spectroscopy.
The onset temperature of the superconducting transition in the in-plane resistivity was approximately 14~K.
A laboratory-built low-temperature scanning tunneling microscope (STM) was used for the STS measurements.
The sample mounted on the STM head was cooled down to 4.2~K in pure helium gas and then cold-cleaved {\it in situ} to expose a clean $ab$ surface.
The measurements were made in a helium gas environment.
Surface topographic images were obtained in constant-current mode, and the tunneling spectra $\mathrm{d}I/\mathrm{d}V$ were acquired by numerical differentiation of the measured $I$--$V$ characteristics at each point within the scan area.

 \begin{figure}
 \includegraphics[width=70mm]{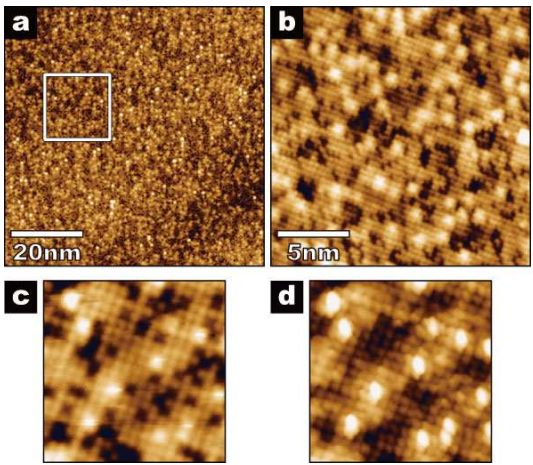}
 \caption{\label{fig1} (Color online)
    (a) Large-scale surface topography of the cleaved surface at 4.2~K ($72 \times 72$~nm$^2$).
    (b) Magnified topography of the region outlined by the box in panel (a), revealing a square atomic lattice with $a_0 \sim 3.8$\AA\ ($V=1.0$~V, $I=40$~pA).
    [(c) and (d)] Tunneling conductance dependence of the STM image in the same region ($6.8 \times 6.8$~nm$^2$ obtained at 18~K) for (c) low conductance (500~mV, 0.5~nA) and (d) high conductance (50~mV, 0.5~nA).
    }
 \end{figure}

Figure \ref{fig1}(a) presents a typical large-scale topographic STM image of the cleaved surface observed at 4.2~K.
The entire image is covered by small patches of bright and dark contrast, similar to previous STM observations.\cite{massee:09075544}
The magnified topography with atomic resolution shown in Fig.~\ref{fig1}(b) reveals a square lattice with a periodicity of approximately 3.8~\AA, which is consistent with the tetragonal lattice constant of the single crystal used in this study.\cite{mizuguchi:jpsj2009}
The atoms observed in the STM image presumably correspond to the apical chalcogen atoms of the Fe(Se,Te)$_4$ tetrahedra located above the Fe plane.
The STM image also exhibits bright spots and dark regions, where the atomic lattice is no longer visible.
The high-conductance (low-bias) STM image shown in Fig.~\ref{fig1}(d) provides clues to the origin of these features.
First, although no atoms are visible in the darkest regions of the low-conductance ($\sim 1$~nS) image [Figs.~\ref{fig1}(b) and \ref{fig1}(c)], atoms appear in the dark regions of the high-conductance ($\sim 10$~nS) image.
Therefore, the areas of dark contrast in the STM images do not correspond to chalcogen vacancies.
Second, the bright spots become sharper in the high-conductance image and thus seem to be particles located on the chalcogen sites.
The number of bright spots is approximately 4.5\% of the number of chalcogen sites, which agrees well with the proportion of excess iron in the crystal.

 \begin{figure}
 \includegraphics[width=75mm]{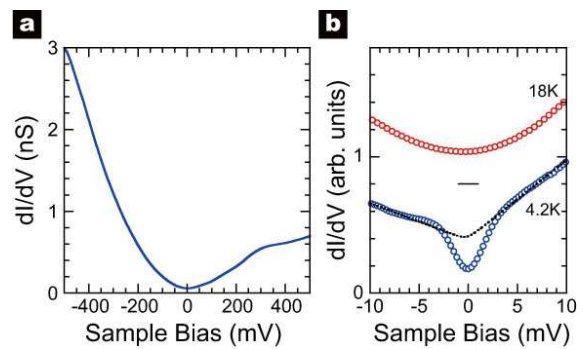}
 \caption{\label{fig2} (Color online)
    (a) Spatially averaged tunneling conductance spectrum in the high-energy range, measured at 4.2~K.
    (b) Low-energy part of the spatially averaged spectrum measured below (4.2~K) and above (18~K) $T_{\mathrm{c}}$, indicating that the gap feature at low temperature corresponds to the superconducting gap.
    The dotted line on the 4.2~K spectrum indicates the calculated V-shaped background conductance (see text).
    The 18~K spectrum has been shifted vertically for clarity.
    }
 \end{figure}

 \begin{figure}[b]
 \includegraphics[width=50mm]{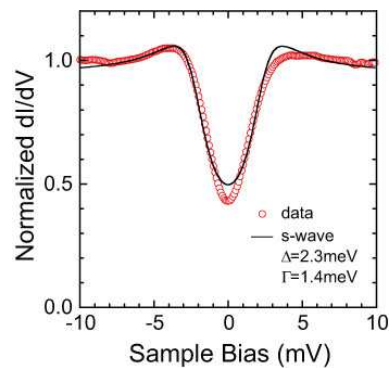}
 \caption{\label{fig3} (Color online)
    Spatially averaged spectrum at 4.2~K normalized to the background conductance represented by the dotted line in Fig.~\ref{fig2}(b).
    The curve indicates a fit of the calculated DOS for an $s$-wave superconductor to the data.
    }
 \end{figure}

The spatially averaged tunneling spectra are presented in Fig.~\ref{fig2} and show the overall features of the DOS in this iron-chalcogenide superconductor.
In the high-energy range [Fig.~\ref{fig2}(a)], the spectrum exhibits an asymmetric V-shaped form with high conductance at the occupied states.
While the occupied DOS increases steeply with increasing negative bias to $-500~$meV, there is a kink at approximately $300$~meV in the unoccupied DOS, representative of a (pseudo)gap feature.
The high-energy spectrum of this iron chalcogenide is very similar to those of the strongly underdoped cuprate superconductors,\cite{kohsaka:science2007} although the two classes of materials have quite different electronic structures.
The high-energy spectra of other iron-based superconductors have not yet been reported; but because most spectra measured by STS exhibit a V-shaped background,\cite{millo:prb2008,boyer:08064400,pan:08080895,yin:prl2009,massee:prb2009,yin:physicac2009} this is probably a common characteristic that is related to the normal state of the iron-based compounds.
In the low-energy part of the spectrum measured at 4.2~K shown in Fig.~\ref{fig2}(b), a dip in the DOS centered at the Fermi energy is superimposed on the V-shaped background, indicating the existence of an energy gap.
The disappearance of this feature above $T_{\mathrm{c}}$ is evidence that this energy gap is responsible for the superconductivity.

In order to determine the value of the superconducting gap $\Delta$, the contribution of the V-shaped background conductance was removed using the following procedure.
First, the background on both sides of the region, displaying the gap feature ($|V|>5$~meV), was extracted from the average spectrum measured at 4.2~K.
The backgrounds for the positive and negative bias regions were separately fitted by linear functions, which were then extended to the Fermi energy.
The resulting V-shaped background was convoluted with the Fermi function at 4.2~K to simulate thermal broadening and is indicated by the dotted line in Fig.~\ref{fig2}(b).
Finally, the measured spectrum was normalized to the simulated V-shaped background and is shown by the circles in Fig.~\ref{fig3}.
The normalized spectrum was fitted using the $s$-wave BCS gap function with a Dynes broadening factor $\Gamma$, as shown by the curve in Fig.~\ref{fig3}.
The resulting value of the gap is $\Delta = 2.3$~meV, which gives the ratio $2\Delta/k_{\mathrm{B}}T_{\mathrm{c}} \sim 3.8$, slightly larger than that for a conventional superconductor.
Because of the relatively high temperature of the measurements with respect to $T_{\mathrm{c}}$ and the arbitrary nature of the background estimation, we were unable to determine the symmetry of the gap from the fitting procedure.
We note that the $d$-wave gap function with $\Delta=3$~meV and $\Gamma=1$~meV gives almost the same curve as the fitted $s$-wave function.

 \begin{figure}
 \includegraphics[width=70mm]{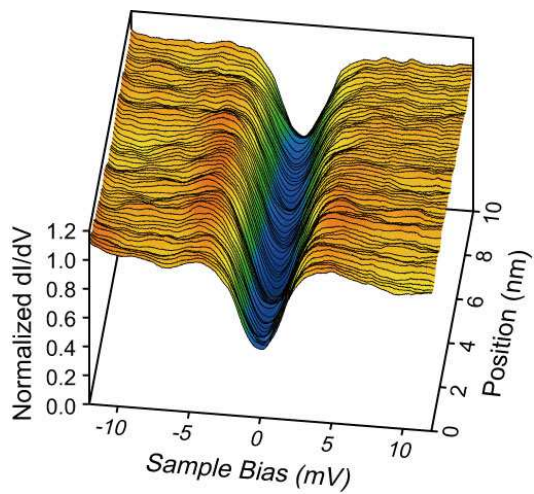}
 \caption{\label{fig4} (Color online)
    Normalized spectra at 4.2~K taken along a 10-nm long linecut across the sample surface.
    All of the raw spectra were normalized to the background conductance shown by the dotted line in Fig.~\ref{fig2}(b).
    }
 \end{figure}

The shape of the gap feature observed in the normalized spectra measured at different points on the sample surface seems to be rather uniform.
Figure~\ref{fig4} displays a series of spectra taken along a 10-nm-long linecut across the surface; each spectrum has been normalized to the background conductance given by the dotted line in Fig.~\ref{fig2}(b).
The magnitude of the gap does not show significant spatial variation, in marked contrast to the cuprate superconductors, which exhibit gross gap inhomogeneity.
The zero-bias conductance and the gap edge peaks vary slightly with location, and the values of $\Delta$ obtained by fitting each normalized spectrum to the $s$-wave function follow a sharp distribution with a small standard deviation of $\sigma_{\Delta} =0.23$~meV.

 \begin{figure}
 \includegraphics[width=70mm]{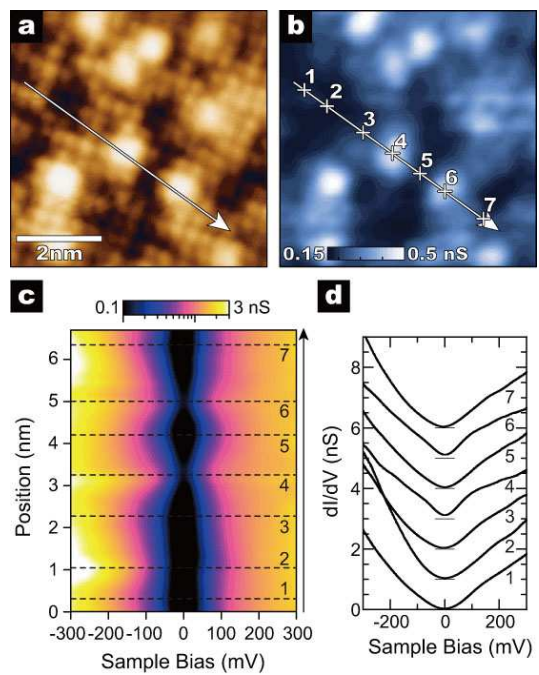}
 \caption{\label{fig5} (Color online)
    Spatial evolution of the high-energy background electronic states measured at 18~K.
    (a) Surface topography obtained at $V=20$~mV and $I=0.4$~nA.
    (b) Differential conductance map at 50~mV for the same area as (a).
    (c) Tunneling spectra obtained along a linecut denoted by the arrows in (a) and (b); intensity is shown by a logarithmic scale.
    (d) Typical spectra, labeled 1--7, extracted from the linecut (shifted vertically for clarity).
    The setup parameters for the STS measurements were $V_{\mathrm{set}}=400$~mV and $I_{\mathrm{set}}=0.4$~nA.
    }
 \end{figure}

While the superconducting gap opens homogeneously below $T_{\mathrm{c}}$, the normal-state electronic states in the high-energy range show spatial variation.
Figure~\ref{fig5} shows a representation of the spatial evolution of the high-energy LDOS measured above $T_{\mathrm{c}}$.
The differential conductance map measured at 50~mV shown in Fig.~\ref{fig5}(b) correlates well with the high-conductance STM image shown in Fig.~\ref{fig5}(a).
Thus, the LDOS spatially varies in accord with the surface topography.
In order to visualize the spectral changes occurring with location, Fig.~\ref{fig5}(c) shows a series of tunneling spectra measured along a linecut denoted by the arrows in Figs.~\ref{fig5}(a) and \ref{fig5}(b).
Seven typical spectra taken at the labeled positions are shown in Fig.~\ref{fig5}(d).
In the darkest region of the STM image (spectrum 2), the LDOS exhibits a simple V-shape with strong bias asymmetry, and the low-energy spectral weight is strongly depressed.
In contrast, at the positions of bright spots in the STM image (spectra 4 and 6), the LDOS around 100~meV is enhanced, and the asymmetry and V-shape are less pronounced, with substantial spectral weight remaining at low energy.
This change in the nature of the LDOS remains up to 1~eV, as can be seen in the high-bias STM image in Fig.~\ref{fig1}(b).
As mentioned earlier, the bright spots are most likely related to excess iron present between the layers.
Therefore, the spatial variation in the normal-state LDOS seems to be caused by the excess iron, although the mechanism by which this occurs is unclear.
The weak spatial variation in the normalized low-energy LDOS in the superconducting state probably originates from this spatially inhomogeneous background electronic structure.

In conclusion, scanning tunneling microscopy and spectroscopy have been performed on the iron chalcogenide superconductor Fe$_{1+\delta}$Se$_{1-x}$Te$_{x}$ both above and below $T_{\mathrm{c}}$.
Atomically resolved topographic STM images of the cleaved surface were acquired, revealing a square lattice without any reconstruction.
The tunneling spectra reveal a superconducting gap with an average value of approximately 2.3~meV.
The normal-state background electronic structure shows strong spatial variation, probably due to the excess iron in the crystal.
Further investigation of the LDOS and superconducting gap in the simplest iron-based compounds will ensure progress in understanding the mechanism involved in the iron-based superconductors.

\end{document}